\documentclass{aa}
\usepackage{amssymb}

\usepackage{graphicx}

\begin{document}

\title{Analytical solutions for energy spectra of electrons
accelerated by nonrelativistic shock-waves in shell type supernova remnants}
\author{V.N. Zirakashvili\inst{1,2}, \and F.Aharonian \inst{2,3}}
\date{Received ; accepted }

\titlerunning{Energy spectra of electrons accelerated in
supernova remnants}

\subtitle{}

\offprints{V.N. Zirakashvili}

\institute{Pushkov Institute of Terrestrial Magnetism, Ionosphere
and Radiowave Propagation, 142190, Troitsk, Moscow Region, Russia\
\and Max-Planck-Institut f\"{u}r\ Kernphysik, D-69029, Heidelberg,
Postfach 103980, Germany\ \and Dublin Institute for Advanced
Studies, 5 Merrion Square, Dublin 2, Ireland }

\abstract
{Recent observations of  hard X-rays and very high energy gamma-rays
from a number of  young shell type supernova remnants indicate
the importance of detailed  quantitative studies of energy spectra of relativistic
electrons formed  via  diffusive shock acceleration
accompanied  by intense nonthermal emission through
synchrotron radiation and inverse Compton scattering.}
{The aim of this work was derivation of exact asymptotic
solutions of the kinetic equation which describes the energy
distribution of shock-accelerated electrons
for an arbitrary  energy-dependence of the  diffusion coefficient.}
{The  asymptotic solutions at low and very high energy domains coupled
with numerical calculations in the intermediate energy range allow
analytical presentations of  energy spectra of electrons for
the entire energy region.}
{Under the assumption that the energy losses of electrons are
dominated by synchrotron cooling, we derived the exact asymptotic
spectra of electrons without any restriction on the
diffusion coefficient. We also obtained
simple analytical approximations which describe,
with accuracy better than ten percent, the  energy spectra of
nonthermal emission of shock-accelerated electrons due to
the synchrotron radiation and inverse Compton scattering.}
{The results can be applied for interpretation of
X-ray and gamma-ray observations of shell type supernova remnants,
as well as other nonthermal high energy source populations
like microquasars and large scale synchrotron jets of
active galactic nuclei.}

\keywords{cosmic rays--acceleration-- galaxies}

\maketitle

\section{Introduction}

The nonthermal X-ray emission detected from a number of young
shell-type SNRs (for a review see e.g. Vink \cite{Vink2006}) is
generally interpreted as synchrotron radiation of relativistic
electrons accelerated diffusively by shock waves to multi-TeV
energies (see e.g. Drury et al. \cite{Druryetal2001}). Two
other (alternative) radiation mechanisms - the bremsstrahlung of
subrelativistic electrons and the inverse Compton (IC) scattering
of  moderately relativistic electrons  - are not sufficiently effective to explain
the observed X-ray fluxes.  The  discovery of TeV gamma-rays from
young SNRs, in particular from Cas A  (Aharonian et al.
\cite{Aharonian2001}), RX J1713.7-3946 (Enomoto et al.
\cite{Enomoto2002}; Aharonian et al. \cite{Aharonian2004}) and  RX
J0852.0-4622  (Katagiri et al \cite{Katagiri2005}; Aharonian et
al. \cite{Aharonian2005}) provides unambiguous evidence of
acceleration of particles (electrons and/or protons) to energies
100 TeV and beyond. Unfortunately even the very high quality
morphological and spectrometric studies of young SNRS in TeV
gamma-rays performed with the HESS array of  imaging atmospheric
Cherenkov telescopes do not allow robust conclusions concerning
the origin of gamma-rays.  For example,   in the case of the best
studied gamma-ray emitting SNR, RX  J1713.7-3946, the hadronic
model  can explain satisfactorily  both the overall energetics
and the spectral features of TeV gamma-ray emission.
Nevertheless,  the inverse Compton origin of gamma-rays
remains an alternative option, provided that the
magnetic field in the gamma-ray production region  does not
exceed $10 \mu \rm G$ (Aharonian  et al. \cite{Aharonian2006}).

Power-law distributions of relativistic particles are readily
formed at astrophysical shocks, in particular in young shell type
SNRs, through  the so-called diffusive shock acceleration (DSA)
mechanism (Krymsky \cite{krymsky77}, Bell \cite{bell78}, Axford et
al. \cite{axford77}). The maximum energy of accelerated particles
and the shape of the spectrum around and beyond the maximum energy
(the so-called "cutoff region") is  determined by the competition
between the acceleration  and escape rates, as well as, in the
case of electrons, by radiative (synchrotron and inverse Compton)
energy losses. In this regard, the most important information
about the accelerator is contained in the energy distribution of
particles in the cutoff region.  The spectrum of relativistic
electrons in the cutoff region is formed under conditions when
particles gain energy during the shock crossing and lose energy
simultaneously. In young SNRs,  the radiative cooling of electrons
results in formation of highest energy tails of synchrotron X-rays
and inverse Compton  gamma-rays, at $\geq 1$ keV and $\geq 10$
TeV, respectively. Thus, the comparison of  hard X-ray and TeV
gamma-ray observations with the spectral features predicted by the
DSA model in these energy bands gives direct information about the
key parameters characterizing the process of particle
acceleration. In this regard,  analytical presentations of the
energy  distributions  of high energy electrons and the spectra of
their synchrotron and IC radiation components provide effective
tools for   studies of nonthermal processes in SNRs.

So far,  analytical solutions for the electron spectra at the
plane  shock have been derived  only for a special case of
\textit{energy-independent} diffusion coefficient (Bulanov \&
Dogiel \cite{bulanov79}, Webb et al. \cite{webb84}, Heavens \&
Meisenheimer \cite{heavens87}).  At the same time, for
typical shock speeds of several thousand km/s
in young SNRs, the particles, in particular electrons, have a chance to be
boosted  to multi-TeV energies only if the acceleration proceeds
in the most effective  way, the so-called Bohm diffusion limit.
This implies strong\textit{
energy-dependence} of the diffusion coefficient, $D(E) \propto E$.
Therefore,  it is important to derive analytical solutions for the
electron spectra  for a more realistic case of diffusion.

In this paper we present exact asymptotic solutions  for the
high-energy tails of  distributions of electrons formed at the
plane shock,  assuming arbitrary energy-dependencies of both the
diffusion coefficient and the energy lose rate of electrons.  The
spectrum of electrons in the low-energy domain where the radiative
cooling can be neglected,  is well known from the solution of the
transport equation characterizing the acceleration at the shock
without losses.  On the other hand, assuming that the cooling of
electrons at very high energies is dominated by synchrotron
losses, we obtained simple analytical expressions for the exact
asymptotic solutions in the high energy domain. In order to extend
the  analytic presentations to a broader  energy range, we  should
"glue"  the asymptotic solutions  applicable at very low and very
high energies, using numerical calculations performed at
intermediate energies. We treat separately the electron
distributions at the shock, as well as downstream and upstream of
the shock, and calculate the spectra of  synchrotron and IC
radiation in these regions. We present these spectra in forms of
simple analytical approximations which appear significantly
different from ones often used in the literature for the fits of
spectral measurements  and subsequent interpretations  of the
X-ray and TeV gamma-ray observations of SNRs in terms of
synchrotron and IC radiation of multi-TeV electrons.

\begin{figure}[t]
\includegraphics[width=8.0cm]{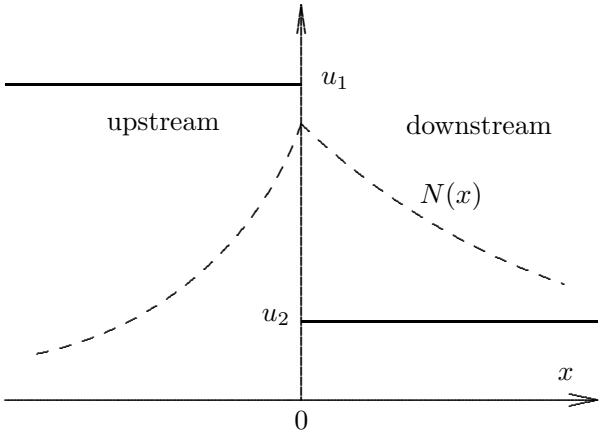}
\caption{The shock speeds  in the
upstream and downstream regions (solid lines) and  the spatial distributions of
electrons (dashed lines), accelerated at the plane shock.  }
\end{figure}

\section{Asymptotic form of the electron spectra}

Below we consider  acceleration of electrons by the plane shock.
The upstream plasma moves with a speed  $u=u_1$ towards the shock
from $-\infty $ of the $x$ axis (see Fig.1). Then, the downstream
speed is $u_2=u_1/\sigma$,  where $\sigma $ is the shock
compression ratio. The electron momentum distribution $N(p,x)$
obey the following equation written in the shock frame upstream
and downstream of the shock:
\begin{equation}
\frac \partial {\partial x}D\frac {\partial N}{\partial x}-
u\frac {\partial N}{\partial x}+\frac 1{p^2}\frac {\partial }{\partial p}
p^2b(p)N=0
\end{equation}
Here $b(p)$  is the energy lose rate of particles ($b  > 0$).  Hereafter we
assume that the synchrotron losses dominate  over the escape of
electrons from the system.  The electron momentum distribution
$N(p)$ is normalized to  $n=4\pi \int p^2dpN(p)$, where $n$ is the
number density of accelerated particles.

Boundary condition at the shock front, $x=0$,  can be written as
\begin{equation}
u_1\frac p{\gamma _s}\frac {\partial N_0}{\partial p}= \left.
D_2\frac {\partial N}{\partial x}\right| _{x=+0}- \left. D_1\frac
{\partial N}{\partial x}\right| _{x=-0} \ ,
\end{equation}
where $\gamma _s=3\sigma /(\sigma -1)$ is the power-law index of
particles accelerated at the absence of  energy losses, $N_0(p)$
is the electron distribution at the shock front, $D_1$ and $D_2$
are the diffusion coefficients upstream and downstream of the
shock, respectively.

Analytical solutions of  Eqs.(1) and (2)  can be obtained  for
simplified conditions, e.g. at  low energies when  the radiative
energy losses can be neglected, or in the case of
energy-independent diffusion-coefficient.  On the other hand, it
is possible to derive asymptotic solutions for an arbitrary
diffusion coefficient in the very high energy regime,  where  the
energy losses of particles dominate over their acceleration.
In this regime, it is
convenient to present the solution for the steady-state energy
distribution of electrons  in the following form:
\begin{equation}
N_{1,2}= K_{1,2}(x,p)\exp[S_{1,2}(x,p)] \ ,
\end{equation}
where indices  $1,2$ correspond to the upstream and downstream
regions, respectively. In the asymptotic regime, when losses
dominate over the acceleration, the function $S(x,p)$ obtains large
negative values. According to the standard asymptotic method we assume
that $S(x,p)$ is proportional to $b(x,p)$. After substitution of
Eq. (3) into Eq. (1) and keeping all  terms which are proportional
to $b^2$ and $b$, one finds

\begin{equation}
D_{1,2}\left( \frac {\partial S_{1,2}}{\partial x}\right) ^2
+b_{1,2}\frac {\partial S_{1,2}}{\partial p}=0
\end{equation}
\[
D_{1,2}K_{1,2}\frac {\partial ^2S_{1,2}}{\partial x^2}+
2D_{1,2}\frac {\partial K_{1,2}}{\partial x} \frac {\partial
S_{1,2}}{\partial x}
\]
\begin{equation}
-u_{1,2}K_{1,2}
\frac {\partial S_{1,2}}{\partial x}+
\frac 1{p^2}\frac {\partial }{\partial p}p^2b_{1,2}K_{1,2}=0 \ .
\end{equation}
Denoting by
\begin{equation}
S_0(p)=\left. S_1(p)\right| _{x=0}=\left. S_2(p)\right| _{x=0} \ ,
\end{equation}
\begin{equation}
K_0(p)=\left. K_1(p)\right| _{x=0}=\left. K_2(p)\right| _{x=0} \ ,
\end{equation}
and  performing the same expansion for the boundary condition
given by  Eq.(2),  we obtain
\begin{equation}
\frac {u_1p}{\gamma _s}\frac {\partial S_0}{\partial p}= {D_2}
\left. \frac {\partial S_2}{\partial x}\right| _{x=0}- {D_1}
\left. \frac {\partial S_1}{\partial x}\right| _{x=0} \ ,
\end{equation}
\begin{equation}
\frac {u_1p}{\gamma _s}\frac {\partial K_0}{\partial p}= {D_2}
\left. \frac {\partial K_2}{\partial x}\right| _{x=0}- {D_1}
\left. \frac {\partial K_1}{\partial x}\right| _{x=0} \ .
\end{equation}

The derivatives of functions $S_1$ and $S_2$ on $x$ and the
derivative $\partial S_0/\partial p$ can be found from Eqs.(4) and
(8):
\begin{equation}
\left. \frac {\partial S_1}{\partial x}\right| _{x=0}= \frac
{\gamma _s}{u_1p}\sqrt{\frac {b_1}{D_1}} \left(
\sqrt{b_1D_1}+\sqrt{b_2D_2}\right),
\end{equation}

\begin{equation}
\left. \frac {\partial S_2}{\partial x}\right| _{x=0}= -\frac
{\gamma _s}{u_1p}\sqrt{\frac {b_2}{D_2}} \left(
\sqrt{b_1D_1}+\sqrt{b_2D_2}\right) \ .
\end{equation}

\begin{equation}
\frac {\partial S_0}{\partial p}=-\frac {\gamma ^2_s}{u_1^2p^2}
\left( \sqrt{b_1D_1}+\sqrt{b_2D_2}\right) ^2
\end{equation}
The  solution of the last equation is
\begin{equation}
S_0=-\frac {\gamma ^2_s}{u^2}\int ^p_0\frac {dp'}{p'^2}
\left( \sqrt{b_1(p')D_1(p')}+\sqrt{b_2(p')D_2(p')}\right)^2.
\end{equation}
Taking the first derivative of Eq.(4) on $x$,  and using Eqs (10)
and (11) we find
\begin{equation}
\left. \frac {\partial ^2S_1}{\partial x^2}\right| _{x=0}= -\frac
{b_1}{2D_1}\frac {\partial }{\partial p}\ln{\sqrt{\frac
{b_1}{p^2D_1}} \left( \sqrt{b_1D_1}+\sqrt{b_2D_2}\right) },
\end{equation}

\begin{equation}
\left. \frac {\partial ^2S_2}{\partial x^2}\right| _{x=0}= -\frac
{b_2}{2D_2}\frac {\partial }{\partial p}\ln{\sqrt{\frac
{b_2}{p^2D_2}} \left( \sqrt{b_1D_1}+\sqrt{b_2D_2}\right) } \ .
\end{equation}

After  substitution of
these  functions  into Eq. (5), and using Eq. (9),  we  arrive at the
following ordinary differential equation
\begin{equation}
p\frac {\partial K_0}{\partial p}K^{-1}_0=-\frac 12+
 \frac {\sqrt{D_2}p\frac {\partial
}{\partial p}\sqrt{b_2} +\sqrt{D_1}p\frac {\partial }{\partial
p}\sqrt{b_1}} {\sqrt{b_1D_1}+\sqrt{b_2D_2}} ,
\end{equation}
the solution of which can be written in the following form:
\begin{equation}
K_0(p)\propto p^{-1/2}\exp{\int dp\frac {\sqrt{D_2}\frac {\partial
}{\partial p}\sqrt{b_2} +\sqrt{D_1}\frac {\partial }{\partial
p}\sqrt{b_1}} {\sqrt{b_1D_1}+\sqrt{b_2D_2}}} \ .
\end{equation}
Eqs. (3),(13) and (17) determine the asymptotic form of the
electron spectrum at the shock front $N_0(p)$ for an arbitrary
diffusion coefficient. In the case of an energy-independent
diffusion, it coincides  with the exact solution  derived by Webb
et al. (\cite{webb84}).

These expressions can be significantly simplified in the case of
the same energy-dependence of the diffusion coefficient and the
energy lose rate  upstream and downstream: $b_2=b=\xi b_1$,
$D_2=D=\kappa D_1$. In this case

\begin{equation}
K_0(p)\propto \sqrt{b/p}   \ ,
\end{equation}
and Eq. (3) is reduced to
\begin{equation}
N_0\propto \sqrt{\frac bp}\exp \left[ -\frac {\gamma ^2_s}{u_1^2}
\left( 1+\frac 1{\sqrt{\xi \kappa }}\right) ^2 \int ^p_0\frac
{dp'}{p'^2}b(p')D(p')\right] \ .
\end{equation}

The integrations of $F_1=\int ^0_{-\infty }dxN(x,p)$ and $F_2=\int
_0^{\infty }dxN(x,p)$   give   the integrated electron spectra
$F_1(p)$ and $F_2(p)$ in the upstream and downstream regions,
respectively. They are of great practical interest in many
astrophysical situations. At large momenta they are
determined by the ratios of the electron spectrum at the shock
front to the absolute values of derivatives given by Eqs.(10) and
(11):
\begin{equation}
F_1(p)=\frac {\xi }{1+\sqrt{\kappa \xi }}\frac {u_1p}{\gamma
_sb(p)}N_0(p) \ ,
\end{equation}
\begin{equation}
F_2(p)=\frac {\sqrt{\kappa \xi }}{1+\sqrt{\kappa \xi }}\frac
{u_1p}{\gamma _sb(p)}N_0(p) \ .
\end{equation}

Eqs. (19)-(21) describe the form of the spectrum at large
energies. At small energies the losses are negligible and the
electron  spectrum is power-law with an index $\gamma _s$. In order
to describe  the transition between this two extreme regimes,
Eq.(1)  has been solved numerically using  an implicit
finite-difference scheme.

\subsection{The case of Bohm diffusion}

Below we consider the most interesting  case which assumes that
the diffusion of electrons proceeds in the  so-called Bohm
diffusion regime and the energy losses of electrons are dominated
by synchrotron radiation. We present the diffusion coefficient  in
the form  $D=\eta cr_g/3$, where $r_g=pc/qB$ is the gyroradius of
particles, $B$ is the magnetic field strength, $q$ and $m$ are the
electric charge and mass of the electron, respectively. The factor
$\eta\geq 1 $  allows deviation of the diffusion coefficient from
its minimum value $\eta =1$ (the nominal Bohm diffusion). The
synchrotron loss rate  averaged over the pitch angles is
$b(p)=4q^4B^2p^2/9m^4c^6$. Since the Bohm diffusion and
synchrotron losses are determined by the magnetic field strength,
the parameter $\xi =\kappa ^{-2}$, where $\kappa $ is the ratio of
the magnetic field upstream to the magnetic field downstream,
$\kappa =B_1/B_2$. Eq. (19)  shows that in the high energy cut-off
region the spectrum has a "Gaussian" type behavior, $N_0 \propto
\exp (-p^2/p^2_0)$, where
\[
\frac {p_0}{mc}=(1+\kappa ^{1/2})^{-1}\frac {mcu_1}{\gamma
_s\sqrt{2\eta q^3B/27}}=
\]
\begin{equation}
\frac {2.86\cdot 10^8}{\gamma _s\eta ^{1/2}(1+\kappa
^{1/2})} \left( \frac {u_1}{3000\ \mathrm{km\ s}^{-1}} \right)
\left( \frac {B}{100\ \mu \mathrm{G} }\right) ^{-1/2} \ .
\end{equation}
Here the shock speed $u_1$ and the magnetic field in the
downstream region $B_2=B$ are normalized to ${3000\ \mathrm{km\
s}^{-1}}$, and $100 \ \mu \mathrm{G}$, which are quite typical for
shells of young SNRs.

At $p=p_0$, the characteristic  lifetime  of electrons  due to  the
synchrotron cooling  is
\[
\tau =p_0/b(p_0)=8.6  \gamma _s(1+\kappa ^{1/2})\eta ^{1/2}\times
\]
\begin{equation}
\left( \frac
{u_1}{3000\ \mathrm{km\ s}^{-1}} \right) ^{-1}\left( \frac {B}{100\ \mu
\mathrm{G} }\right) ^{-3/2} \ \mathrm{yr}\ .
\end{equation}
The synchrotron losses have a strong impact on formation of
the energy spectrum of electrons at energies for which this time
is small compared to the characteristic dynamical times of  the
source, e.g.  the age of the accelerator. If this condition is
fulfilled, the electrons with such energies are concentrated in
the shock vicinity,  and therefore the plane shock approximation is well
justified.

In the case of  standard shock acceleration in the Bohm diffusion
regime  with $\kappa =1$ and $\gamma _s=4$,  we combined  the
asymptotic analytical solutions, obtained at low and very high
energies,  with numerical calculations performed in the transition
region. This allows  a simple analytical presentation of the
electron spectrum at the shock  over the entire energy range:
\begin{equation}
N_0(p)\propto p^{-4}[1+0.66(p/p_0)^{5/2}]^{9/5}\exp
(-p^2/p^2_0) \ .
\end{equation}
The spectrum of electrons given by Eq.(24) is shown in Fig. 2. At
low and large energies  Eq. (24) coincides with the exact
asymptotic solutions:  $N_0(p) \propto p^{-4}$ and  $N_0(p)
\propto p^{1/2} \exp(-p^2/p^2_0)$, respectively. In the cutoff
region, the spectrum  is proportional to the product of two terms
--  a power-law term with a positive slop ($p^{1/2}$) and  an
exponential term, $\exp(-p^2/p^2_0)$. While  the first term can be
interpreted as a pile-up\footnote{The possible appearance of a
pile-ups in the electron spectra within the so-called "box" model
of shock acceleration has been noticed by Drury et al.
(\cite{Drury99}).},  it in fact does not sow-up because  the
second (exponential) term effectively cancels this feature. As a
result,  an almost perfect power-law ($p^{-4}$  type) spectrum is
formed up to $p \sim p_0$, with a super-exponential cutoff
afterwards (see Fig.2).

It is easy to relate, using Eq. (1)  and boundary condition given by Eq. (2)
the integrated spectra in
downstream, upstream and the spectrum at the shock:

\begin{equation}
F_2+\kappa ^2F_1=l\frac {p_0}{p}N_0(p), \ \mathrm{and} \ l=\frac
{u_1p^2}{\gamma _sbp_0}.
\end{equation}
 This relation is valid for all energies. It
implies  that for $\kappa =1$ (i.e. equal magnetic field strengths
upstream and downstream)  the integrated spectrum $F=F_1+F_2
\propto N_0p^{-1}$.

The case of different magnetic fields in the
downstream and upstream regions,
more specifically, for the scenario when  the magnetic field downstream
is stronger than magnetic field upstream by a factor of $\kappa
^{-1} =\sqrt{11}$, is of practical interest, in particular
for nonthermal emission of young supernova remnants.   This case,
which  corresponds to the increase of the  isotropic random
B-field at the shock with the "standard" compression factor
$\sigma =4$, can be described by the following analytical
approximations for the electron distributions at the shock front:
\begin{equation}
N_0(p)\propto (p/p_0)^{-4}[1+0.523(p/p_0)^{9/4}]^{2}\exp
(-p^2/p^2_0) \ ,
\end{equation}
and in the upstream region
\begin{equation}
F_1(p)\propto 0.70l(p/p_0)^{-3}[1+1.7(p/p_0)^{3}]^{5/6}\exp
(-p^2/p^2_0) \ .
\end{equation}
For the downstream,  the integrated energy spectrum of electrons can
be found from Eq. (25):

\begin{equation}
F_2(p)=N_0(p)l\frac {p_0}{p}-0.09F_1(p) \ .
\end{equation}

Note that the spectrum integrated over the upstream region, at
small energies  is very flat, $F_1 \propto p N_0(p) \propto p^{-3}
$. This has a simple explanation  related to the spatial scale of
the electron distribution; in the upstream region it is
proportional to the diffusion coefficient.

The integrated spectra of electrons in the upstream and downstream
regions given by Eqs. (26)-(28) as well as results obtained
numerically are shown in Fig.3.

The dependence of the spectra given by Eqs. (24), (26)-(28)
on the factor $\eta $ is expressed through the momentum $p_0$.

\section{Synchrotron  radiation}

The synchrotron emissivity of electrons
is determined  as
\begin{equation}
\epsilon (\omega )=\frac {\sqrt{3}Bq^3}{2\pi mc^2}
\int p^2dpN(p)R(\omega /\omega _c)
\end{equation}
where $\omega _c$ is the characteristic frequency of synchrotron
radiation $\omega _c=1.5\ qBp^2/m^3c^3$. The function $R$
describes synchrotron radiation of a single electron in magnetic
field with chaotic directions.  Crusius \& Schlickeiser
(\cite{crusius86}) derived an exact expression for  $R(\omega
/\omega _c)$ in terms of Whittaker's function. With an accuracy of
several percent $R(\omega /\omega _c)$ can be presented in a
simple analytical form
\begin{equation}
R(\omega /\omega _c)= \frac {1.81\exp (-\omega /\omega _c)}
{\sqrt{(\omega _c/\omega )^{2/3}+(3.62/\pi )^2}} \ .
\end{equation}

These functions are compared in Fig.4.

 The energy flux
($J(\omega)=\omega {\rm d}N_{\mathrm{rad}}/{\rm d}\omega$) of
synchrotron radiation produced at the shock is determined by the
integral of the emissivity along the line of sight $l$ and over
the solid angle  $\Omega $ $J(\omega )=\int dld\Omega
\epsilon $. This means that it is determined by the integrated
spectrum $F(p)$.

The integration of  Eq. (29), using Eqs. (24) and (30),
results in the energy flux of radiation $J(\omega )=\int
dpa(p,\omega )\exp (-g(p,\omega ))$, where $g(p,\omega )=\omega
/\omega _c+p^2/p^2_0$,  $a(p,\omega )\propto p^2F(p)R(\omega
/\omega _c)\exp (g(p,\omega ))$. At large frequencies, when the
argument of the exponent is large this integral is mostly
determined by the momentum interval around the momentum $p_*$
which minimizes the function $g(p,\omega )$. Thus at large
frequencies the energy flux of radiation can be written as
$J(\omega )=a(p_*,\omega )\exp (-g(p_*,\omega ))\sqrt{2\pi
/g''(p_*,\omega )}$. Here $g''$ is the second derivative of $g$ on
$p$. The momentum $p_*$ is given by the expression:
\[
\frac {p_*}{mc}=\left( 9\frac {\omega m^3c^3u_1^2}{\eta \gamma
_s^2q^4B^2}\right) ^{1/4}\frac {1}{\sqrt {1+\kappa ^{1/2}}}=8.3
\times 10^7 \times
\]
\begin{equation}
\frac { (\hbar \omega /1 \rm  \ keV)^{1/4}} {\gamma _s^{1/2}  \eta
^{1/4} \sqrt {1+\kappa ^{1/2}} } \left( \frac {u_1}{3000
\mathrm{km/s}} \right)^{1/2} \left(\frac {B}{100\ \mu \mathrm{G}
}\right)^{-1/2} \ .
\end{equation}
Note that  at the fixed $\omega$, the characteristic
frequency of the synchrotron radiation $\omega_c $ corresponding
to  $p_*$,  is smaller than $\omega$.

In the regime of Bohm diffusion the energy
flux of synchrotron radiation has an asymptotic form
\begin{equation}
J \propto \omega ^{3/8}\exp (-\sqrt{\omega /\omega _0}) \ ,
\end{equation}
where
\begin{equation}
\omega _0=\frac {81}{16\left( 1+\kappa ^{1/2}\right) ^2} \frac
{u_1^2mc}{\eta \gamma ^2_sq^2} \ ,
\end{equation}
or for the  corresponding energy of  synchrotron photons
\begin{equation}
\epsilon _0= \hbar \omega _0= \frac {\mathrm{2.2\ keV}}{\eta
(1+\kappa ^{1/2})^2}\left( \frac {u_1}{\mathrm{3000\ km\ }s^{-1}}
\right) ^2\frac {16}{\gamma _s^2} \ .
\end{equation}

At small energies the integrated downstream synchrotron spectrum
is proportional to $\omega ^{-{\frac {\gamma _s-2}{2}}}$. In order to
describe analytically the  synchrotron spectrum over
the entire energy range, which should  coincide  with the  exact
asymptotic spectra  at low and high energies,  we use
numerical calculations  performed in the intermediate energy range. In
particular, in the  case of $\gamma _s=4$ and $\kappa
=1$, we suggest  the following approximate formula
\begin{equation}
J(\omega )\propto \frac {\omega _0}{\omega }\left[ 1+ 0.46\left(
\frac {\omega }{\omega _0} \right) ^{0.6}\right] ^{11/4.8} \exp
(-\sqrt{\omega /\omega _0} \ ) \ ,
\end{equation}
which provides an accuracy  better than 10 percent.

The energy spectrum of electrons at the shock $N_0(p)$ (see Eqs.
(24)), is shown in Fig.2. Because the cooling length of electrons
downstream the shock is inversely proportional to the momentum,
the spatially integrated electron spectrum appears steeper, in
particular, in the case of $\kappa=1$, $F(p)=F_1(p)+F_2(p) \propto
N_0(p)/p$. Note that below the cutoff region $F(p) \propto
p^{-5}$. The corresponding spectrum of synchrotron radiation of
these electrons is shown in Fig.2. At small frequencies, $J(\omega
)\propto \omega ^{-1}$.

The synchrotron spectrum obtained using the so-called $\delta
$-function approximation is also shown in Fig.2. It is assumed,
that the synchrotron emission of the  electron is concentrated at
the frequency $\omega =0.29\omega _c$. This corresponds to the
function $R(\omega /\omega _c)\propto \delta (0.29-\omega /\omega
_c)$. Obviously this approximation does not provide a satisfactory fit.


\begin{figure}[t]
\includegraphics[width=8.5cm]{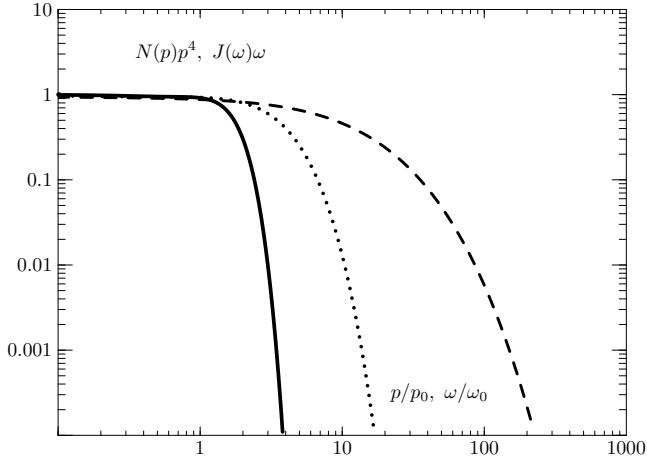}
\caption{The electron distribution at the shock front (solid line)
and the spatially integrated spectrum of synchrotron radiation
(dashed line) produced by  electrons  accelerated at the shock
with compression ratio $\sigma =4$ and equal upstream and
downstream magnetic fields ($\kappa =1$). The spectrum of
synchrotron radiation obtained using the $\delta $-function
approximation is also shown (dotted line). }
\end{figure}

In the  case of $\gamma _s=4$, $\kappa =1/\sqrt{11}$ we
found the following spectra produced upstream and downstream the
shock:
\begin{equation}
J_1(\omega )\propto 0.175\left[ 1+ 1.27\left( \frac {\omega
}{\omega _0}\right)^{3/4}  \right]^{1/2} \exp \left
(-\sqrt{3.32\frac {\omega }{\omega _0}}  \right) ,
\end{equation}
\begin{equation}
J_2(\omega )\propto \frac {\omega _0}{\omega }\left[ 1+
0.38\sqrt{\frac {\omega }{\omega _0}}  \right]^{11/4} \exp
\left( -\sqrt{\frac {\omega }{\omega _0}}  \right) .
\end{equation}
Note that the spectrum of synchrotron radiation produced upstream the shock at small
frequencies is very flat, $J(\omega)=\mathrm{const}$, which
reflects the  hard electron spectrum  in that region (see Eq.(27) and Fig.3).

The  spectra of synchrotron radiation in the upstream and
downstream regions  given by Eqs.(36) and (37) are shown in Fig.5. 
The synchrotron flux in the downstream region is significantly
enhanced because of the  larger (compressed)  magnetic field.
The corresponding spectra obtained numerically are also shown.

We should note that the exponential terms in Eqs. (35),(36) and
(37) are relatively slow functions (proportional to $\exp
[-(\omega/\omega_0)^{1/2}]$). Moreover,  these exponential terms
are multiplied to  power-law terms $(\omega/\omega_0)^s$ with
positive indices, $s \ge 0$,  which effectively compensate the
exponential terms at  frequencies $\omega \sim \omega_0$. For that
reason the parameter $\omega_0$  only formally can be  treated
as a cutoff  frequency.  In reality,  the breaks (cutoffs) in
the synchrotron spectra  appear at much higher frequencies,
namely at $\omega \geq 10 \omega_0$.

\begin{figure}[t]
\includegraphics[width=8.5cm]{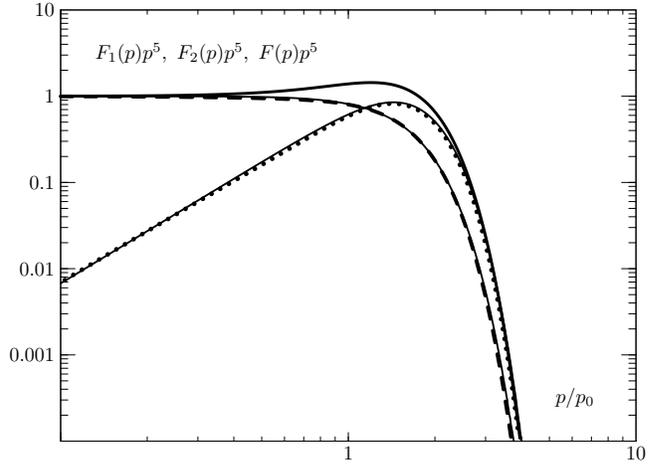}
\caption{The integrated upstream ($F_1$; dotted line) and
downstream ($F_2$; dashed line) electron spectra given by Eqs.
(27), (28), as well as the overall $F=F_1+F_2$ spectrum (solid
line) for the shock with compression ratio $\sigma =4$ and the
ratio of the magnetic fields downstream and upstream
$k^{-1}=\sqrt{11}$. The spectra $F_1$ and $F_2$ obtained
numerically are also shown (thin solid lines). }
\end{figure}

\begin{figure}[t]
\includegraphics[width=8.5cm]{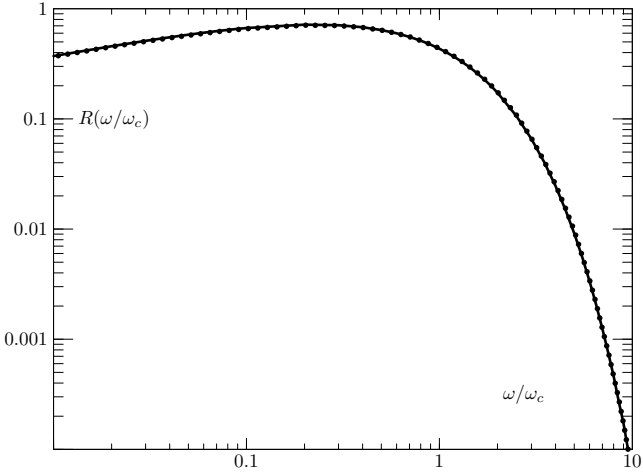}
\caption{Comparison of the function $R(\omega /\omega _c)$
tabulated by Crusius and Schlickeiser (\cite{crusius86}) (dotted
line) and the function given by the approximate expression (30)
(solid line).  }
\end{figure}

\begin{figure}[t]
\includegraphics[width=8.5cm]{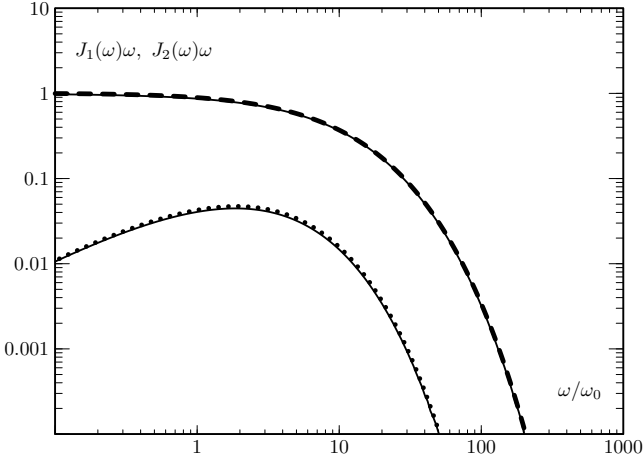}
\caption{The energy spectra  of synchrotron radiation given by
Eqs. (36),(37) and produced by electrons  in downstream (dashed
line) and upstream (dotted line) regions of the shock with
compression ratio $\sigma =4$ and the ratio of the magnetic fields
downstream and upstream $\kappa ^{-1}=\sqrt{11}$. The
corresponding spectra obtained numerically are also shown (solid
lines).}
\end{figure}

The integrated downstream electron spectrum $F_2$ in Fig.3 and the
corresponding synchrotron spectrum $J_2$ in Fig.5 may be applied
to the supernova remnants with ages  lager than the cooling time given by
Eq.(23), i.e.  when the electron spectrum in the cut-off region
is determined by the energy losses of
electrons. Electrons with energies below cut-off energy are
accelerated with power-law spectrum and  advected downstream. It
was also assumed that these electrons lose all their energy
downstream that is of course valid for energies larger than the
energy of the electron which has the  cooling time comparable with
the age of the remnant. For smaller energies a three-dimensional
geometry of the shock wave should be taken into account.  Consequently,
the integrated downstream electron spectrum and the
corresponding  synchrotron radiation spectrum  appear  flatter
compared to the  ones shown in Figs.3,5.

\subsection{The radiation  zones}

The asymptotic solutions derived in Sec.2 can be used for study of
coordinate-dependence of electron distributions. Near the shock
front, the spatial distribution of electrons  has an exponential
dependence.  The inverse exponential lengths in the upstream and
downstream regions are given by Eqs. (10) and (11). Since the
synchrotron radiation is produced mainly by electrons with
momentum $p_*$, the exponential lengths  of emissivity (29) in the
upstream $l_u$ and downstream $l_d$ regions of the shock can be
found from Eqs. (10),(11) and (31):

\begin{equation}
l_d=\frac {9}{4\sqrt{3(1+\kappa ^{1/2})}} \left( \frac {
m^9c^{17}u_1^2\eta }{\omega q^{12}B^6\gamma _s^2}\right) ^{1/4} \
,
\end{equation}
and
\begin{equation}
l_u=l_d\kappa ^{-5/4} \ .
\end{equation}
In particular, for $\gamma _s=4$ we have
\[
l_d=\frac {1.4\cdot 10^{17}\ \mathrm{\ cm}}{\sqrt{1+\kappa
^{1/2}}} \eta ^{1/4}\left( \frac {u_1}{\mathrm{3000\ km\ }s^{-1}}
\right) ^{1/2}\times
\]
\begin{equation}
\left( \frac B{100\ \mu \mathrm{G}}\right) ^{-3/2} \left( \frac
{\hbar \ \omega }{\mathrm{1\ keV\ }} \right) ^{-1/4} \ ,
\end{equation}
for the radiation width in downstream, and a factor of  $\kappa
^{-5/4}$ larger in upstream. Note that the above  equations are
correct only for  large frequencies, $\omega >\omega _0$.

The radiation width in downstream (40) is relatively small. This
permits to explain the  narrow X-ray filaments observed in SNRs
and to estimate the magnetic field strength (see e.g. Berezhko et
al. \cite{berezhko02}, V\"olk et al. \cite{voelk05}).

\section{Inverse Compton scattering}

In the isotropic photon field,  the emissivity of the inverse
Compton  radiation of an electron with  Lorentz factor $\gamma$  is
described as  (Blumenthal \& Gould \cite{blumenthal70})
\[
\epsilon _{\mathrm{IC}}(\omega )=2\hbar cr_e^2\int \limits ^\infty _{\frac
{\omega }{4\gamma ^2\left( 1-\frac {\hbar \omega }{\gamma
mc^2}\right)}} d\omega 'n(\omega ') \frac {\omega }{4\gamma
^2\omega '} \left[ 1-2\frac {\omega ^2}{\omega _c^2}+\right.
\]
\begin{equation}
\left. \frac {\omega }{\omega _c }\left( 1+2\ln \frac
{\omega }{\omega _c}\right)
+\frac {\hbar ^2\omega ^2}{2m^2c^4\gamma ^2
\left( 1-\frac {\hbar \omega }{\gamma mc^2}\right)}\left( 1-\frac {\omega
}{\omega _c}\right) \right] .
\end{equation}
Here $\omega _c=4\omega '\gamma ^2\left( 1-\frac {\hbar \omega
}{\gamma mc^2}\right) $ , $r_e=q^2/mc^2$ is the classical electron
radius and $n(\omega)$ is the energy distribution of  the target
photons. Hereafter we assume that the target photons are described by
a Planckian  distribution
\begin{equation}
n(\omega )=\frac {\omega ^2/(c^3\pi ^2)}{\exp (\hbar \omega
/\kappa_{\mathrm{B}}T)-1} \ ,
\end{equation}
where $T$ is the temperature of radiation {and $\kappa
_{\mathrm{B}}$ is the Boltzman constant}. It is easy to show,
using Eqs. (41) and (42),  that   at large frequencies $\omega
>> \gamma ^2\kappa _{\mathrm{B}}T/\hbar $ the IC  emissivity in the Thompson regime
behaves as   $\epsilon _{\mathrm{IC}}(\omega) \propto \omega \exp
(-\hbar \omega /4\gamma ^2\kappa _{\mathrm{B}}T)$. At very high
energies  the integration over the electron energy distribution at
the shock and the use of the method described in the Section
3,  give the following asymptotic form  of the IC spectrum in the
Thompson limit:
\begin{equation}
J^{\mathrm{IC}}\propto \omega ^{7/8}\exp (-\sqrt{\omega /\omega _b}) \ ,
\end{equation}
with
\begin{equation}
\omega _b=\frac {27}{2\eta \left( 1+\kappa^{1/2}\right) ^2} \frac
{\kappa _{\mathrm{B}}Tu_1^2m^2c^2}{\hbar \gamma ^2_sq^3B} \ ,
\end{equation}
or, in terms of energy of gamma-rays,
$\epsilon _b=\hbar \omega _b$, we have:
\begin{equation}
\epsilon _b=\frac {\mathrm{1.2\ TeV}}{\eta
(1+\kappa^{1/2})^2}\left( \frac {u_1}{\mathrm{3000\ km\ }s^{-1}}
\right) ^2 \frac {100\ \mu \mathrm{G}}{B}\frac {T}{2.7\
\mathrm{K}} \frac {16}{\gamma _s^2} .
\end{equation}
As before, it is  assumed that synchrotron losses dominate over IC losses,
i.e.  the energy density of the background radiation
$U_{\mathrm{rad}}<<B^2/8\pi $. In the case of 2.7 K CMBR this
implies $B \ge  3 \mu \rm G$.

Using the exact asymptotic spectra at low and high energies,
and  the results of numerical calculations at  intermediate energies,
we can describe the broad-band energy flux of the
inverse Compton  (Thompson) scattering  by simple analytical
expressions.   In particular, for $\kappa=1$ one has
\begin{equation}
J^{\mathrm{IC}}(\omega )\propto \frac {\omega _b}{\omega }\left[ 1+
0.36\left( \frac {\omega }{\omega _b} \right) ^{0.7}\right]
^{15/5.6} \exp \left(-\sqrt{\frac {\omega }{\omega _b}} \ \right)
\end{equation}

\begin{figure}[t]
\includegraphics[width=8.5cm]{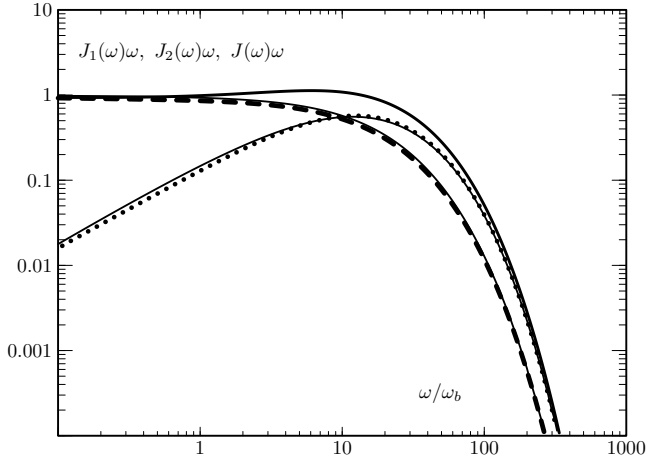}
\caption{The spectrum of IC radiation given by Eqs. (47),(48) and
produced by electrons accelerated downstream (dashed line) and
upstream (dotted line) the shock with compression ratio $\sigma
=4$ and the ratio of the magnetic strength downstream and upstream
$\sqrt{11}$. The sum
$J^{\mathrm{IC}}=J^{\mathrm{IC}}_1+J^{\mathrm{IC}}_2$ is shown by
the solid line. It is assumed that IC scattering proceeds in the
Thompson limit,  $b_{\rm KN}=0$, and that synchrotron losses
dominate over the IC loses. The corresponding spectra obtained
numerically are also shown (thin solid lines).}
\end{figure}

For the "standard" scenario of shock acceleration
with  $\gamma _s=4$ and  $\kappa=1/\sqrt{11}$,  we obtain  the
following IC spectra produced upstream and downstream
the shock:
\begin{equation}
J^{\mathrm{IC}}_1(\omega )\propto 0.20\left[ 1+ 0.75\left( \frac {\omega
}{\omega _b}\right) ^{7/8} \right] \exp \left (-\sqrt{\frac
{\omega }{\omega _b}} \ \right) ,
\end{equation}
\begin{equation}
J^{\mathrm{IC}}_2(\omega )\propto \frac {\omega _b}{\omega }\left[ 1+
0.31\left( \frac {\omega }{\omega _b}\right) ^{0.6} \right]
^{25/8} \exp \left( -\sqrt{\frac {\omega }{\omega _b}} \ \right) .
\end{equation}
The spectra given by Eqs. (47) and (48) are shown in Fig.6. They are
derived under the assumption that the Compton
scattering occurs  in the Thompson regime. However,  in many cases
one  should take into account the Klein-Nishina effect. The
importance of this effect in the case of black-body radiation
field is determined by the  parameter
\[
b_{KN}=4\frac {\kappa _{\mathrm{B}}Tp_0}{m^2c^3}= \frac
{0.52}{\gamma _s\eta ^{1/2}(1+\kappa^{1/2})} \times
\]
\begin{equation}
\left( \frac
{u_1}{3000\ \mathrm{km\ s}^{-1}} \right) \left( \frac {B}{100\ \mu
\mathrm{G} }\right) ^{-1/2}\frac {T}{2.7\ \mathrm{K}} \ .
\end{equation}
The Thompson limit corresponds to $b_{KN}<<1$, while the
Klein-Nishina effect becomes significant  already at $b_{KN} \geq
0.1$.  The IC spectra of electrons calculated for different values
of the parameter $b_{KN}$ are shown in Fig.7.

\begin{figure}[t]
\includegraphics[width=9.0cm]{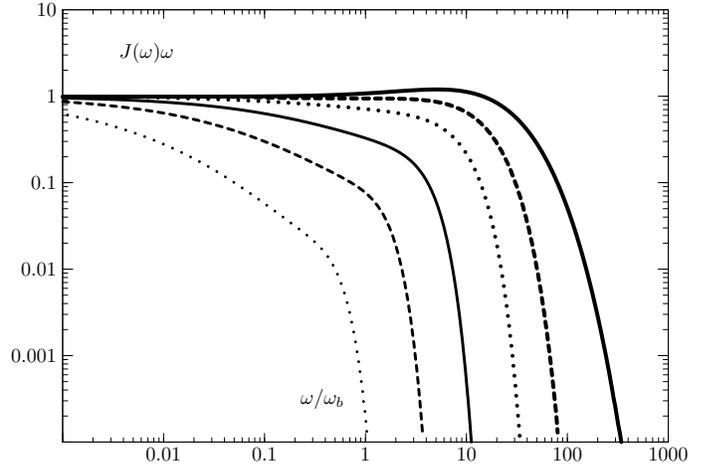}
\caption{The spectra of IC radiation produced by electrons
accelerated at the shock with compression ratio $\sigma =4$ and
the ratio of the magnetic strength downstream and upstream
$\sqrt{11}$ for different values of $b_{KN}$: $b_{KN}=0$ (solid
line), $b_{KN}=0.1$ (dashed line), $b_{KN}=0.3$ (dotted line),
$b_{KN}=1.0$ (thin solid line), $b_{KN}=3.0$ (thing dashed line)
and $b_{KN}=10.0$ (thin dotted line). The case of dominating
synchrotron losses is considered. }
\end{figure}

As long as the energy losses of electrons are dominated by
synchrotron radiation, the  Klein-Nishina effect  makes steeper
the spectrum of IC gamma-rays.
In particular,  it moves the region of the spectral cutoff to lower
energies.  Below the cutoff region,  the energy spectrum
of  IC gamma-rays  still can be described by a  (quasi) power-law.
The Klein-Nishina effect changes the slope of the gamma-ray
spectrum  significantly,  even for relatively small values of the
parameter $b_{KN}$.

\section{Discussion}

The detection of nonthermal X-ray emission of synchrotron origin
clearly demonstrates the existence of multi-TeV electrons in
different astrophysical source populations like shell type SNRs --
pulsar wind nebulae, microquasars,  small and large scale jets in AGN,  \textit{etc.}
Given the severe synchrotron  losses of electrons, which increase with
energy as $dE/dt \propto E^2$, the acceleration efficiency
in these objects  should be very high in order to boost the electrons
to energies well beyond 1 TeV.
In particular, in young SNRs with  typical shock speeds
$u_1 \sim 2000-3000 \ \rm km/s$, which are established
as prominant nonthermal X-ray emitters,  the diffusive shock
acceleration of electrons should proceed in the regime close to
the Bohm diffusion limit,  otherwise the synchrotron spectrum  would
break before achieving the X-ray domain.

Although the basic concepts
of diffusive shock acceleration are  deeply studied and well understood
(see e.g. Malkov and Drury \cite{malkov01}), many  important
details remain   unexplored. This concerns, in particular the radiation features
of the shock accelerated electrons.  Over  the recent years,
several numerical calculations of high energy radiation by
particles accelerated in young SNRs have been performed
with emphasis on the non-linear effects caused by protons and
nuclei on  the shock structure (see e.g.
Ellison et al. \cite{Ellison2000}, Berezhko and V\"olk \cite{berezhko04}).

The shock accelerated electron spectra have been studied analytically
(e.g. by  Webb et al. \cite{webb84}), but  only for the
case of energy-independent  diffusion coefficient. Therefore
the results of these early works  generally cannot be applied to the most interesting
sources with synchrotron X-ray emission.  These sources, as noticed above,
require diffusive shock acceleration of electrons in the regime
close to the Bohm diffusion which is characterized by a
diffusion coefficient  $D(E) \propto E$.

The aim of this paper  was  a fully analytical treatment of  the energy
spectra  of shock-accelerated electrons and their  synchrotron and inverse
Compton radiation for  an arbitrary  energy-dependent
diffusion coefficient.  In particular, under the assumption
of dominance of synchrotron losses, we derived exact  asymptotic
solutions in the  high  energy  region.  For the energy-dependent diffusion
coefficient written in a rather general form $D\propto p^\alpha$,
the electron spectrum contains an exponential  term
$\exp [-(p/p_0)^\beta]$ with $\beta=\alpha+1$ (see Eq.(19)).
Thus,  only in the  case of energy-independent diffusion ($\alpha=0$) the
electron spectrum is characterized by an exponential cutoff. In the
most interesting case of  Bohm diffusion,
the spectral cut-off is of Gaussian type,  $\exp [-(p/p_0)^2]$.
The corresponding cutoff in the spectrum of syncrotron radiation
appears much smoother (see also Fritz
\cite{fritz89} for discussion of this effect).  It is described
by an exponential term
$\exp[-(\omega/\omega_0)^{\beta/(2+\beta)}]
=\exp [-(\omega/\omega_0)^{(\alpha +1)/(\alpha +3)}]$.
For the energy independent diffusion coefficient we have
$J(\omega) \propto \exp[-(\omega/\omega_0)^{1/3}]$,
while in the case of Bohm diffusion
$J(\omega) \propto \exp[-(\omega/\omega_0)^{1/2}]$.

We should note that the  $\exp [-(\omega/\omega_0)^{1/2}]$ type behavior
of the synchrotron spectrum in the cutoff region
\textit{formally} is similar to the one derived
for the  generally  assumed  "power-law with
exponential cutoff" electron spectrum,
$N_0(p) \propto p^{-\Gamma} \exp(-p/p_0)$,
and using the $\delta $-function approximation for
calculation of the synchrotron radiation
(see e.g. Reynolds \cite{Reynolds98}).  However, this is
simply a coincidence.  First of all, the $\delta$-functional
approximation leads to a wrong synchrotron spectrum;
the exponential cut-off in the electron spectrum
results in the synchrotron spectrum
with a $\exp (-\omega ^{1/3})$ type cut-off.
More importantly, the spectra  of shock accelerated electrons
given by  Eq.(26)-(28)
are quite different from the simplified  "power-law with
exponential cutoff"  assumption. Correspondingly, because of the
dependence $\omega_c \propto p^2 B$, the effect on
the spectrum  of  synchrotron radiation appears
even more pronounced (see Fig.2).

Exact  solutions for electrons are possible only at low and high
energies, namely in the energy intervals where (\textit{i}) the
radiative losses of electrons are negligible,  and  (\textit{ii})
the radiative energy loss rate exceeds the acceleration rate. The
indices  of the  power-law factors in the asymptotic solutions
(19),(32) and (43)  significantly  differ from the power-law
indices of the electrons, as well as the synchrotron and IC
spectra below the cut-off regions.  In order to connect these
solutions we performed numerical  calculations in the intermediate
energy intervals. This  allowed us to obtain, within accuracy
better than 10 percent, simple analytical presentations for  the
spectra  of electrons and their synchrotron and inverse Compton
radiation components.

Eqs. (35)-(37) and Eqs. (46)-(48) properly  describe the  relative
contributions of radiation from  the regions upstream and
downstream the shock. At the same time,  they do not allow  direct
calculations of  the absolute synchrotron and inverse Compton
luminosities. However, the normalisation constants  can be easily
calculated  from the total energy release through the synchrotron
and inverse Compton channels.

It should be  noted  that in this paper the momentum $p_0$ in the
electron spectrum, and the corresponding frequencies $\omega _0$
and $\omega _b$ which characterize the positions of cut-offs in
the spectra of synchrotron and IC radiation components, are found
from the exact asymptotic solutions of kinetic equations.
Interestingly, the results appear quite close, within a factor of
2,  to   the estimates  derived from the condition "acceleration
rate = synchrotron cooling rate". However, as it follows  from
Fig. 2-6, the real cut-offs  appear, in fact, at much higher
frequencies.  In the case of synchrotron radiation, the spectrum
continues without an indication of a break or a cutoff up to  $20
\omega _0$. The spectrum of  inverse Compton gamma-rays turns down
also at $\omega \sim 20 \omega _b$ (in the case of constant
magnetic field), or it continues up to $\sim 50 \omega _b$ (in the
case of different magnetic fields upstream
 and downstream).  The more effective extension of
the hard spectrum of IC radiation  is explained by  the larger
spatial scale of electron distribution upstream the shock. This is
not  important  for the synchrotron radiation the contribution of
which from the upstream region is suppressed by an order of
magnitude.  On the other hand, because of the homogeneous
distribution of the target radiation field, 2.7 K CMBR, the IC
gamma-ray contribution at highest energies from  the upstream
region dominates over the contribution from the downstream region.

The results obtained in this paper can be used to fit the  X-ray
spectra observed from young supernova remnants. The value of the
parameter $\omega _0$ found from the fitting procedure should be
compared with the theoretical value given by Eq.(33). For the
known shock velocity of a given supernova remnant, this would
allow an important estimate of deviation of the diffusion regime
from the extreme Bohm limit. The dependence of the energy spectra
 on the factor $\eta $ is expressed via parameters $p_0$,
$\omega _0$, $\omega _b$. If the width of X-ray
filaments is known, the magnetic field strength may be found from
Eq. (40). All this in turn would provide reliable estimates of the
maximum energy of protons and nuclei accelerated in supernova
remnants.

Finally, we  note  that the results of this paper
are applicable for  nonrelativistic shocks with
velocities   up to  $u_1 \sim  (0.1-0.3) c$. The derived
analytical approximations do not  take into account
possible nonlinear effects which
can somewhat modify the spectra of electrons at low and high energies.

\acknowledgement{VNZ acknowledges the hospitality of the
Max-Planck-Institut f\"ur Kernphysik, where this work was carried
out. We thank the anonymous referee for many valuable comments. }

\end{document}